# Polarization squeezing at the audio frequency band for the Rubidium D$_1$ line

XIN WEN,[1, 2]   YASHUAI HAN,[1, 2]   JINYU LIU,[1, 2]   JUN HE,[1, 2, 3]
AND   JUNMIN WANG [1, 2, 3, *]

[1] *Institute of Opto-Electronics, Shanxi University, Tai Yuan 030006, P. R. China;*
[2] *State Key Laboratory of Quantum Optics and Quantum Optics Devices, Shanxi University, Tai Yuan 030006, P. R. China;*
[3] *Collaborative Innovation Center of Extreme Optics, Shanxi University, Tai Yuan 030006, P. R. China*
*\* email: wwjjmm@sxu.edu.cn*

**Abstract**: A 2.8-dB polarization squeezing of the Stokes operator $\hat{S}_2$ for the rubidium D$_1$ line (795 nm) is achieved, with the lowest squeezing band at an audio frequency of 2.6 kHz. Two methods are applied for improving the squeezing towards low frequencies: an orthogonal-polarized locking beam that precludes residual unwanted interference and quantum noise locking method that stabilizes the relative phase between the squeezed vacuum and the local oscillator. The squeezing level is limited by absorption-induced losses at short wavelengths, here is 397.5 nm. The generated polarization squeezed light can be used in a quantum enhanced magnetometer to increase the measurement sensitivity.

## 1. Introduction

Squeezed states are prevalent in the field of quantum physics, since their first generation in 1985 [1]. With their advanced quantum features, squeezed states have been demonstrated in several applications such as quantum teleportation [2], quantum information networks [3, 4], quantum memory [5, 6], and in quantum metrology and precise measurements [7, 8], which are of interest to us. Early experiments produced squeezing at megahertz frequencies, benefiting from the quiet noise background of the laser sources. However, in certain situations, kilohertz squeezing bands are required; for example, in gravitational wave detection (10 Hz–10 kHz) [9-11] and interaction with the atomic media [12, 13]. Squeezed states at the kilohertz frequencies are required for coherent delay and storage in an electromagnetically induced transparency model, owing to the narrow transparent windows [14]. Practical applications in bio-magnetism demand magnetic field measurements at kilohertz and below, such in magneto-cardiography and magneto-encephalography, from a human heart or brain [15], respectively. In the audio frequency regime, squeezing is easily submerged in roll-up noises and the measured squeezing level is limited. Two factors limit the obtained squeezing: the technical noise induced in the detection [16, 17] and squeezing degradation by the noise coupling of the control beams [18, 19]. Several previous researches have revealed that parasitic interference, beam jitter, noise coupling from the local oscillator, and the imbalance in the electronic circuits are the main noise sources in squeezing measurements. In addition, the optical noises in the pump, probe, and lock beams may be induced or transmitted to the squeezed light, causing degradation.

Bowen *et al*. [20] and Schnabel *et al*. [21] placed optical parametric amplifiers (OPA) in a Mach-Zehnder interferometer; such noise cancellation configurations enabled the generation of squeezed vacuums, down to 220 kHz and 80 kHz, respectively. Laurat *et al.* [22] designed a single type-II, self-phase-locked, frequency degenerate optical parametric oscillator (OPO) and the generated two-mode squeezed state reached 50 kHz. McKenzie *et al.* [18] developed a noise dither locking technique to obtain a squeezed vacuum from 280 Hz. In their system, the OPO cavity length was not actively stabilized; hence, long-term running was impossible. The lowest squeezing achieved till date is 1 Hz [19], with two frequency-shifted control lights utilized to sense the cavity length and pump phase, respectively. This scheme eliminated parasitic interference successfully, but two separate lasers had to be used and the system was complex. Oelker *et al.* [23] produced a frequency-dependent rotated squeezing at 1.2 kHz with a filter cavity; it is built to match the gravitational-wave-detection scheme.

The results listed above were presented at a gravitational-wave-detection wavelength of 1064 nm; in order to interact with the atomic media, wavelengths resonant with the atomic transition lines need to be selected. For the rubidium $D_1$ line, as the wavelength of 795 nm is considerably shorter than 1064 nm, problems of absorption and heating arise. The reason is that the 397.5 nm pump beam is close to the lower limit of the transparent window (350-4400 nm) of the PPKTP crystal. Absorption induces additional losses, leading to the degradation of squeezing level. Heating causes instability of the cavity locking, this limits the squeezing band towards low frequencies. Several experimental attempts were made for generating polarization squeezing at 795 nm. Wolfgramm *et al.* [13] obtained a squeezing of 3.6 dB at 1 MHz with the squeezing band ranging from 0.08–2 MHz; Wu *et al.* [24] achieved a polarization squeezing of 4 dB at an analytical frequency of 3 MHz; whereas, Hétet *et al.* [25] obtained a squeezing of 5.2 dB, covering a frequency range, 150–500 kHz. Our group achieved a polarization squeezing of 5.6 dB at 2 MHz, with a frequency band ranging from 0.2–10 MHz [26]. However, these results are all in the high frequency range, far away from the audio frequencies, where additional noise coupling needs to be considered.

In this work, benefiting from a homemade low-noise homodyne detector at audio frequencies and the improvement in the technical noises, we present a 2.8-dB polarization squeezed light, down to 2.6 kHz at the rubidium $D_1$ line. The combination of the following two methods is necessary for low frequency squeezing: The utilization of a frequency-shifted polarization-perpendicular beam, which counter propagates in the cavity, to lock the cavity length and a quantum noise locking method to stabilize the relative phase between the generated squeezed vacuum and the local oscillator (LO). With both the cavity and the LO phase locked, the generated polarization squeezed light can be subsequently used for precise measurements.

## 2.  Noise coupling of the OPO cavity

The two variances of the output field from the OPO cavity can be modeled as a combination of various noise sources [18]:

$$V^{\pm}(\omega) = \left\{ \begin{array}{l} C_s V_s^{\pm}(\omega) + C_l V_l^{\pm}(\omega) + C_\upsilon^{\pm}(\omega) V_\upsilon^{\pm}(\omega) \\ + \alpha^2 \left[ C_p V_p^{\pm}(\omega) + C_\Delta^{\pm} V_\Delta(\omega) \right] \end{array} \right\} / \left| D^{\pm}(\omega) \right|^2 \qquad (1)$$

where $V_s$, and $V_p$ are the fluctuations from the seed and pump field, respectively, $V_l$ is from the intracavity loss, $V_\upsilon$ is the vacuum fluctuation from the output coupler, and $V_\Delta$ is owing to the cavity detuning, which is due to mechanical instability. $\alpha$ is the intracavity fundamental field and is determined by the seed power. The coupling coefficients are as follows:

$$D^{\pm}(\omega) = i\omega + \kappa_a + \begin{bmatrix} 3 \\ 1 \end{bmatrix} \varepsilon^2 \alpha^2 / (2\kappa_b) \mp \varepsilon\beta,$$

$$C_s = 4\kappa_{in}^a \kappa_{out}^a, C_l = 4\kappa_l^a \kappa_{out}^a,$$

$$C_\upsilon^{\pm}(\omega) = \left|2\kappa_{out}^a - D\pm(\omega)\right|^2, C_p = 4\kappa_{out}^a \kappa_{in}^b (\varepsilon/\kappa_b)^2 \tag{2}$$

$$C_\Delta^{\pm} = 8\kappa_{out}^a \begin{bmatrix} 0 \\ 1 \end{bmatrix},$$

where, the parameters, $\kappa_{out}^a, \kappa_{in}^a$, and $\kappa_l^a$ are the decay rates of $\alpha$ induced by the output coupler, input coupler and the loss. $\kappa_{in}^b$ is the decay rate of the intracavity second harmonic field, $\beta$ is owing to the input coupler, and $\kappa_a$ and $\kappa_b$ are the total decay rates for $\alpha$ and $\beta$, respectively. $\varepsilon$ is a nonlinear coupling parameter and $\omega$ is a small frequency shift, relative to the carrier frequency.

As depicted in Eq. (1), the field strength parameter, $\alpha$, couples the pump field and detuning fluctuations and scales with a squared value; hence, it contributes more to the output variances. When the seed field is a vacuum, $\alpha$ is zero; thus, the influence of the pump field and the detuning fluctuations are eliminated. Therefore, we use an OPO in our experiment to isolate the noise source from the intracavity field, eliminating the squared noise sources.

## 3. Improvements for squeezing towards the audio frequencies

The experimental setup is displayed in Fig. 1. A low-noise continuous wave Ti: Sapphire laser works as the fundamental source. The wavelength of the laser is tuned to 794.975 nm, resonating with the rubidium $D_1$ line. A 30-dB optical isolator is used to avoid optical feedback. A 3.6-MHz modulation is applied to a phase-type electric optical modulator (EOM), which is used in the Pound-Drever-Hall scheme [27] for locking cavities. Two bow-tie type cavities accomplish the second harmonic generation (SHG) and the OPO successively and a third cavity, the mode cleaner (MC), is built to optimize the spatial mode of the LO. All the cavities have lengths of approximately 600 mm and a distance of approximately 120 mm between the two concave mirrors. The radii of curvature of the concave mirrors are 100 mm. A 10-mm-long periodically poled KTiOPO$_4$ (PPKTP) crystal is placed at the center of the concave mirrors and the resulting beam waist in it is approximately 40 μm. The PPKTP crystals are placed in copper-made ovens and have temperatures that are actively stabilized to approximately 53° C, for optimization. For the OPO cavity, the transmissivity of the output coupler is 11.5% at 795 nm, whereas the other three mirrors are highly reflective for the fundamental wavelength of 795 nm; the concave mirrors are highly transmissive at 397.5 nm; hence, the pump beam has a single pass through the nonlinear crystal, avoiding the extra loss owing to the absorption of ultra-violet light. The intra-cavity loss is estimated to be 0.4%, resulting in an escape efficiency of 96.6%. A homemade homodyne detector is utilized, with a common-mode rejection ratio (CMRR) greater than 45 dB. It is designed for use at audio frequencies of typically, 10 Hz–400 kHz [28]. The electric dark noise is 16 dB below the shot noise limit and the quantum efficiency of the photodiode (First Sensor, Model: PC20-7) is 95%.

For squeezing at audio frequencies, two techniques are utilized in our improved system. A counter-propagating and orthogonal-polarized beam is used for locking the cavity length of the OPO, avoiding the mixing of extra noise. Quantum noise locking enables the determination of the relative phase between the squeezed vacuum and the LO. In addition, the low noise detector and the optical isolator, after the OPO, provide firm support for squeezing detection. The following paragraphs describe them in detail.

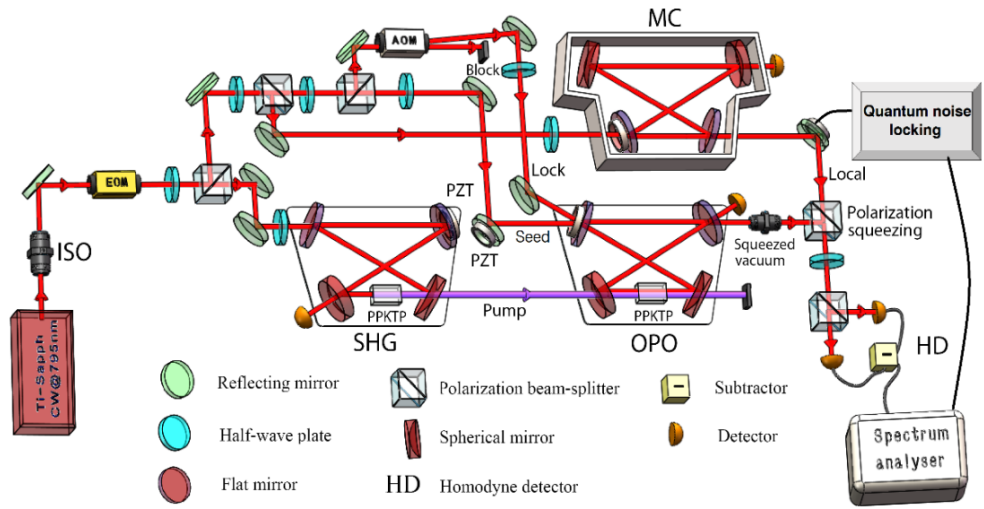

Fig. 1. Experimental setup for polarization squeezing.

First, we use a counter-propagating and orthogonal-polarized beam to lock the OPO cavity. Initially, the lock beam was in a diverse direction only and worked well in the megahertz regime. However, in the audio frequency band, the residual reflection at the crystal surface will couple into the squeezed field, degrading the squeezing level. To overcome this problem, the polarization of the lock beam was changed to a p-polarization, which does not sense the OPO process. For differences between the reflective indices of the polarizations, the resonant frequency differs. In order to ensure the operation of the OPO, the cavity should be locked at the OPO resonant wavelength; hence, we use an acoustic-optical modulator to shift the frequency of the p-polarization beam to approximatley 140 MHz. However, a major heating source arises due to the absorption of the ultra-violet pump beam in the PPKTP crystal, which changes the cavity length; hence, the frequency of the p-polarization beam should be shifted by another 20 MHz. In addition, owing to the variation in the surrounding temperature, the frequency needs to be adjusted slightly to match with the OPO condition.

The other method used in our experiment is quantum noise locking [29], which is critical in obtaining squeezing at a low frequency. It is used to lock the squeezing phase. In an OPO configuration, a squeezed vacuum is obtained; however, the interference visibility of the squeezed and LO fields is poor; thus, the traditional phase locking method cannot be used. The configuration used is illustrated in Fig. 2. Quantum noise locking involves three steps: bandpass filtering, envelope detection, and modulation-demodulation. The settings of the resolution bandwidth (RBW) and the video bandwidth (VBW) on the spectrum analyzer (SA) select the bandwidth and the envelope detection period amplifies the real signal, within a certain band. Then, the output signal is demodulated with a lock-in amplifier and the resultant error signal is fed back to a piezo-electric actuator. The noise locking method depends on the asymmetry of the phase-sensitive variances, and the stability depends on the level of squeezing and anti-squeezing.

In the final step, an optical isolator is placed immediately outside the OPO cavity, precluding light from being reflected back into it, from the detector surface, for instance. On one hand, light may break the locking stability of the cavity, whereas on the other, when reflected again by mirror surfaces, the following parasitic interference may contaminate the squeezing level at low frequencies. .

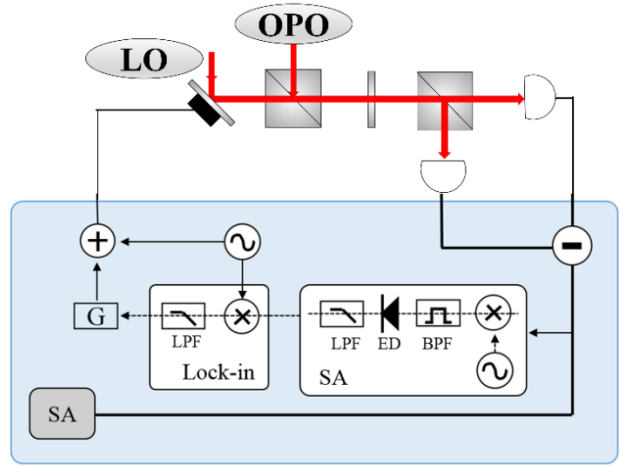

Fig. 2. Noise locking schematic; the electric part is in the blue box.

## 4. Experimental results and discussions

After SHG, approximatley 50 mW of 397.5-nm laser is injected into the second cavity to pump the OPO. The PPKTP crystal in the OPO process is temperature controlled to 52.85 °C; and then the p-polarized locking beam is frequency shifted to match the s-polarized OPO beam. Further, the polarimeter, consisting of a half-wave plate and a polarization beam splitter (PBS), is used to measure the polarization squeezing. The optical axis of the half-wave plate is oriented at 22.5° to either the horizontal or vertical plane. After the polarimeter, the polarized beams are split into two parts and sent to the homodyne detector; the difference signal is used to measure the polarization squeezing. We first measure the squeezing traces in a scanning mode by varying the relative phase between the squeezed vacuum and the LO. The results shown in Fig. 3 are at the center frequencies of 2 MHz and 50 kHz respectively. At 2 MHz, a polarization squeezing of -5.6 dB is obtained and the anti-squeezing is +7.0 dB; whereas at 50 kHz, the squeezing and anti-squeezing levels are -2.3 dB and +6.6 dB, respectively. With a quantum efficiency of 95%, escape efficiency of 96.6%, propagation efficiency of 99%, and an interference visibility of 99.7%, the expected squeezing level at 2 MHz is -6.9 dB [26]. It is clear that the squeezing traces are more stable at high frequencies. However, at low frequencies, fluctuations in the optical path result in the instability of the relative phase between the squeezed vacuum and the LO beam; thus, the noise signal dithers and cannot always stay at the lowest point.

    To form a stable noise spectrum, we use the noise locking method. The generated squeezed vacuum is interfered with a coherent LO beam, monitored on the homodyne detector. We extract the homodyne signal from the alternating output and send it to an SA (Agilent, Model: E4405B). The SA is set to zero span. The center frequency is 2 MHz, the RBW is 300 kHz, VBW is 30 kHz, and the sweep time is 1 s. The output of the SA is demodulated in a lock-in amplifier (SRS, Model: SR830). The modulation by the lock-in amplifier has a frequency of 35.01 kHz and an amplitude of 1.950 V. The error signal is fed back to the piezo-electric transducer (PZT) on the LO beam. By switching the phase in the proportional-integral-derivative controller, either a squeezed phase or an anti-squeezed phase can be achieved. The results are shown in Fig. 4.

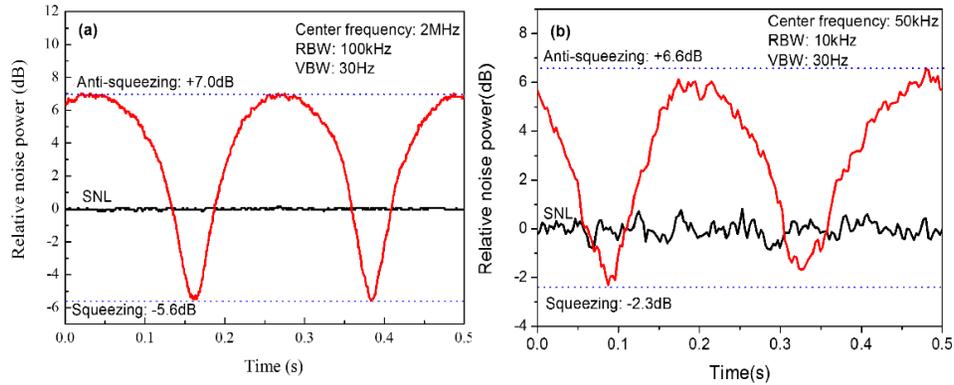

Fig. 3. Measured polarization noise power, as the phase of the local oscillator beam is scanned. (a) Result at an analyzing frequency of 2 MHz, zero-span mode, 100 kHz RBW, and 30 Hz VBW; (b) Result at an analyzing frequency of 50 kHz, zero-span mode, 10 kHz RBW, and 30 Hz VBW. The shot noise level (SNL) trace shows the noise power with the polarized local oscillator beam, but not with polarization squeezing. It gives the noise power level reference at 0 dB. In figure 3(a), -5.6 dB polarization squeezing and + 7.0 dB anti-squeezing are obtained at an analyzing frequency of 2 MHz. In the figure 3(b), -2.3 dB polarization squeezing and + 6.7 dB anti-squeezing are obtained at an analyzing frequency of 50 kHz. Obviously, the result at the low analyzing frequency is noisier than that at the higher analyzing frequency.

The noise traces in Fig. 4 are recorded by a second SA (Agilent, Model: 4396B). For measuring from at 1 kHz–100 kHz, the RBW is 100 Hz and the VBW is 10 Hz; in the frequency range, 2.2 kHz–3 kHz, the RBW is 10 Hz and the VBW is 1 Hz; all the noise traces are averaged 16 times. A flat noise spectrum down to 2.6 kHz is obtained and the measured squeezing level is approximately 2.8 dB. The electric noise is subtracted from the data. The peaks at approximately 19 kHz and 38 kHz are the resonant frequency of the PZT and its second harmonic, respectively; the peak at 35 kHz is the modulation frequency of the lock-in amplifier and that at 88 kHz may be owing to the extra noise in the electric circuits. Compared to the scanning mode, it is obvious that the noise locking method has a good stability.

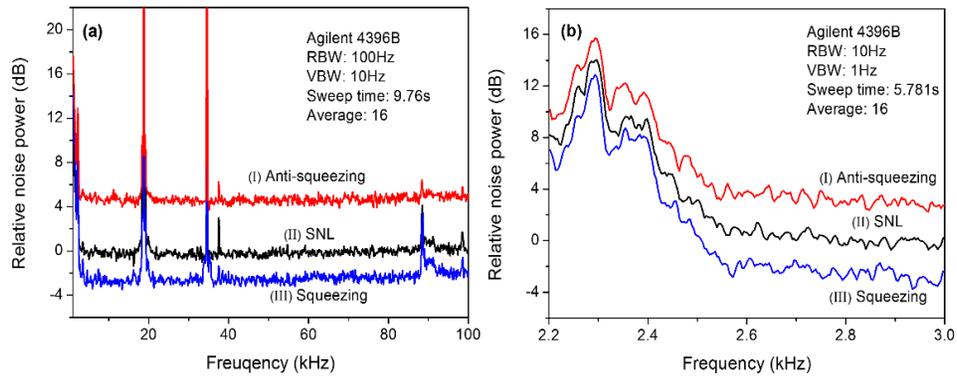

Fig. 4. Quantum noise locking results. In the frequency ranges, (a) 1 kHz–100 kHz; (b) 2.2 kHz–3 kHz. (I), (II), and (III) are the anti-squeezing, shot noise limit, and the squeezing traces, respectively. The squeezing level is approximatley 2.8 dB and the flat noise spectrum, down to 2.6 kHz, is shown clearly.

The polarization squeezing is modeled using Stokes operators and can be illustrated on a quantum Poincaré sphere [30, 31]. Position is given by the mean value and sizes in different directions are determined by the corresponding fluctuations. They are used to express the polarization of lights with quantum features. The polarization squeezed light is composed of two orthogonal lights. Hence, it is possible to get more than one squeezed Stokes operators with both the fields squeezed. Furthermore, instead of the homodyne method, using the polarization quasiprobability function could make all the Stokes operators squeezed simultaneously [32]. However, our polarization squeezing is synthetized by a bright coherent LO and a squeezed vacuum. Thus, only one of the Stokes operators could be squeezed [30, 31]. With the polarimeter model established here, the Stokes operator $\hat{S}_2$ is detected and locked to the squeezing phase. Fig. 5(a) illustrated a quantum Poincaré sphere. The red ball in it represents a certain polarization state. Fig. 5(b) is the visualization of the polarization squeezed state produced in our experiment with the Stokes operator $\hat{S}_2$ squeezed and $\hat{S}_3$ anti-squeezed.

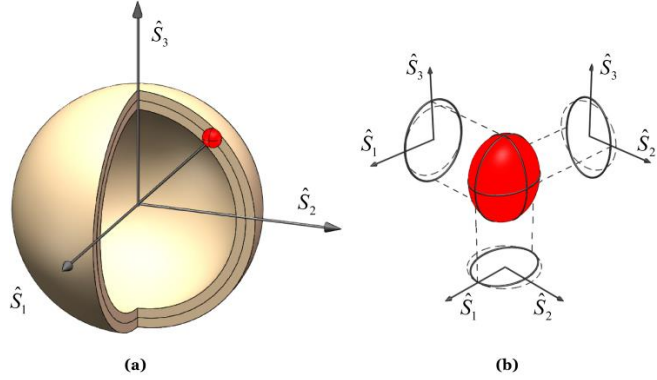

Fig. 5. (a) Illustration of the quantum Poincaré sphere. The red ball on it represents a certain polarization state. Position is given by the mean value and sizes are determined by the fluctuations. (b) Diagrammatic illustration of the Stokes operators, in our experiment. The red ellipsoid is the squeezed noise ball and the three ellipses are the projections at each plane. The dashed circle represents the noise of the coherent state and the solid one shows the squeezing results. The graph clearly shows that the $\hat{S}_2$ operator is squeezed, the $\hat{S}_3$ operator is anti-squeezed, and that the $\hat{S}_1$ operator is the same as the coherent classical state.

The power of the polarization squeezed light is 2 mW, which is limited by the saturated power of the homodyne detector. It is mainly dependent on the coherent LO beam power; as the squeezed vacuum has an almost vacuum intensity, it can be ignored. The degree of polarization of the generated beam is approximatley 3000:1, determined by the PBS used to synthetize the polarization squeezing.

Although commercial homodyne detectors are available, for good gain and stability in balanced detection, we utilize a homemade one. Its quiet noise background at the measurement bands, high CMRR, and special design for the use of photodiodes with high quantum efficiency provide a solid foundation for our squeezing measurements.

As demonstrated by the results above, with the aid of the low noise homodyne detector and the vacuum injected OPO, we got a polarization squeezing down to 2.6 kHz. The combination of the two technique methods is a must to maintain the OPO operation. In the locking beam, both the opposite propagation direction and the perpendicular polarization preclude the noise coupling of the reflected light at the crystal surface and the possible interference. The quantum noise locking method is the only method for locking the phase of an OPO without an additional reference light and the squeezed vacuum is necessary to achieve squeezing at low frequencies. Besides, the optical isolator outside the OPO cavity prevents reflection into the cavity and is beneficial against the parasitic interference.

Compared with the lowest squeezing frequencies, the squeezed band ends at 2.6 kHz still meets some restrictions. One of the reasons is the intrinsic noise of Ti: Sapphire laser, a feedback circuit may be used to stabilize the fluctuations. Second is the roll-up of the detector electric noise, which limits extension to the lower frequencies. Thus, the homodyne detector needs to be improved. Next is the imperfect mechanical stability of our bow-tie type OPO cavities. As for the severe absorption and heating problems, the standing-wave cavities with better stability are inappropriate in the short wavelengths, such as 795 nm. Hence, stabilizing the system by increasing the compactness, enclosing to avoid disturbances from the surroundings and actively isolating the fluctuations, may be beneficial.

## 5. Conclusions

We have used two methods to achieve polarization squeezing at an audio frequency band: an orthogonal-polarized, frequency-shifted locking beam and the quantum noise locking method. A flat noise spectrum as low as 2.6 kHz was obtained with a squeezing level of 2.8 dB. It is difficult to further extend the squeezing band to lower frequencies, where more severe requirements for vibration isolation or noise reduction are expected. Enhancing the system stability and careful control of the cavity loss may enable an increase in the squeezing level. After locking the relative phase between the squeezed vacuum and the LO beam, the LO beam obtains a considerably quieter noise background at its orthogonal port. In future, we intend to use this polarization squeezed light as the probe beam in a magnetic field sensor. The weak magnetic signal will emerge from the squeezed noise spectrum, achieving an enhanced signal-to-noise ratio and increasing the measurement sensitivity.

## Funding